\documentstyle[colordvi,prl,aps,psfig,multicol]{revtex}

\begin{document}

\title{Transmission Phase of an Isolated Coulomb--Blockade Resonance}

\author{H. A. Weidenm\"uller}

\address{Max-Planck-Institut f\"ur Kernphysik, D-69029 Heidelberg,
  Germany}

\date{\today}

\maketitle

\begin{abstract} In two recent papers, O. Entin--Wohlman {\it et
    al.} studied the question: ``Which physical information is carried
  by the transmission phase through a quantum dot?'' In the present
  paper, this question is answered for an islolated Coulomb--blockade
  resonance and within a theoretical model which is more closely
  patterned after the geometry of the actual experiment by Schuster
  {\it et al.} than is the model of O. Entin--Wohlman {\it et al.}. We
  conclude that whenever the number of leads coupled to the
  Aharanov--Bohm interferometer is larger than two, and the total
  number of channels is sufficiently large, the transmission phase
  does reflect the Breit--Wigner behavior of the resonance phase shift.

\end{abstract}

\begin{multicols}{2}

\section{Introduction}
\label{int}

\begin{figure}[h]
 \phantom{.}\vspace{0.1cm}
  \begin{center}
    \leavevmode
    \parbox{0.6\textwidth}
           {\psfig{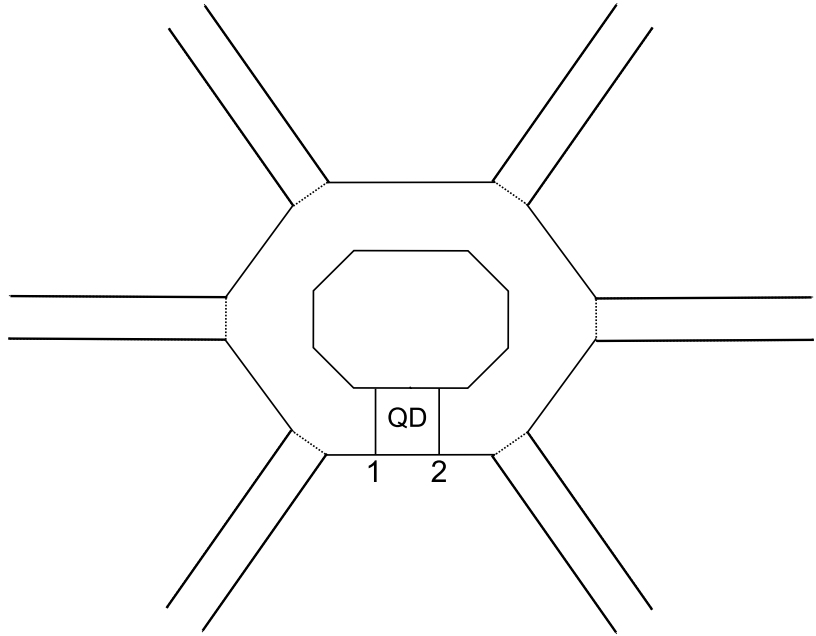}}
  \end{center}
\protect\caption{An AB interferometer containing a QD in one of its
  arms and attached to six external leads (schematic). The barriers
  labelled $1$ and $2$ separate the QD from the AB ring, the dashed
  lines the AB ring from the leads. Barriers and dashed lines are
  considered closed in the geometry defining the Hamiltonian $H_0$ in
  Eq.~(\ref{eq2}).}
\label{fig1}
\end{figure}

In 1997, Schuster {\it et al.}~\cite{sch97} reported on a measurement
of the transmission phase through a quantum dot (QD). These authors
used an Aharanov--Bohm (AB) interferometer with the QD embedded in one
of its arms. The device is schematically shown in Figure~\ref{fig1}.
The current through the device is made up of coherent contributions
from both arms and is, therefore, a periodic function of the magnetic
flux $\phi$ through the AB interferometer. A sequence of
Coulomb--blockade resonances in the QD was swept by adjusting the
plunger gate voltage $V_g$ on the QD. (The plunger gate is not shown
in the Figure). The phase shift $\delta \phi$ of the oscillatory part
of the current was measured as a function of $V_g$. We refer to this
phase shift as to the transmission phase through the QD. As expected,
$\delta \phi$ showed an increase by $\pi$ over each Coulomb--blockade
resonance.

The AB interferometer of Ref.~\cite{sch97} was attached to six
external leads. The complexity of this arrangement was caused by the
failure of an earlier two--lead experiment~\cite{yac95} also aimed at
measuring $\delta \phi$. Instead of a smooth rise by $\pi$ over the
width of each resonance, a sudden jump by $\pi$ near each resonance
was observed. This feature was later understood to be caused by a
special symmetry property of the two--lead experiment~\cite{bue86}:
The conductance $g(\phi)$ and, therefore, the current are symmetric
functions of $\phi$ and, hence, even in $\phi$. Thus, $g$ is a
function of $\cos ( \phi )$ only, and the apparent phase jump of
$\delta \phi$ by $\pi$ is actually due to the vanishing at some value
of $V_g$ near resonance of the coefficient multiplying $\cos ( \phi
)$.

\end{multicols}

\begin{multicols}{2}

Following the work of Ref.~\cite{sch97}, theoretical attention was
largely focused on the least expected feature of the data of
Ref.~\cite{sch97}: A sequence of Coulomb--blockade resonances
displayed very similar behavior regarding the dependence of both, the
conductance and the transmission phase, on $V_g$. In particular,
$\delta \phi$ displayed a rapid drop between neighboring
Coulomb--blockade resonances. For references, see Ref.~\cite{bal99}.
It is only recently that Entin--Wohlman, Aharony, Imry, Levinson, and
Schiller~\cite{ent01,ent02} drew attention to the behavior of $\delta
\phi$ at a {\it single} resonance. In Ref.~\cite{ent01} the first
four authors consider a three--lead situation. Two leads are attached
to the AB interferometer in a fashion analogous to Figure~\ref{fig1}.
The third lead connects directly to the QD. The authors take the arms
of the AB interferometer and the three external leads as
one--dimensional wires. They consider an isolated resonance due to a
single state on the QD. Solving this model analytically, they come up
with a disturbing result: The transmission phase increases by $\pi$
over an energy interval given by that part of the width of the QD
which is due to its coupling to the third lead. As this coupling is
gradually turned off, the rise by $\pi$ of $\delta \phi$ becomes ever
more steep, and eventually becomes a phase jump by $\pi$ as the
coupling to the third lead vanishes. In Ref.~\cite{ent02}, the model
is extended to include additional one--dimensional wires directly
attached to the AB ring and coupled to it in a special way. Again, it
is found that the transmission phase reflects the resonance phase
shift only ``for specific ways of opening the system''\cite{ent02}.
These results immediately poses the following questions: What is the
behavior of $\delta \phi$ for a single resonance in the six--lead case
and, more generally, for any number of leads in a geometry like the
one shown in Figure~\ref{fig1}? Does the rise of $\delta \phi$ by
$\pi$ occur over an energy interval given by the actual width of the
resonance or only by part of that width? And what happens to $\delta
\phi$ as the number of leads is gradually reduced to two? It is the
purpose of the present paper to answer these questions within a
theoretical framework which is more closely patterned after the
experimental situation than is the work of Ref.~\cite{ent01}. In
particular, our work differs from that of Refs.~\cite{ent01,ent02} in
the following respects: We do not assume that the QD is directly
coupled to a lead, we do not make any specific assumptions about the
way in which the leads are coupled to the AB ring, and we allow for an
arbitrary number of leads (as long as this number is at least equal to
two) and of channels in each lead.

\section{Model}

In one of the first theoretical papers~\cite{hack97} addressing the
data of Ref~\cite{sch97}, the transmission phase $\delta \phi$ was
calculated in the framework of a model designed in Ref.~\cite{hack96},
and for the geometry shown in Figure~\ref{fig1}. Inspection of the
curves published in Ref.~\cite{hack97} shows that the transmission
phase rises roughly by $\pi$ over an energy interval roughly equal to
the width of each Coulomb--blockade resonance. The curves shown in
Ref.~\cite{hack97} were, however, calculated in the framework of
specific assumptions on a number of parameters and do not, therefore,
constitute a general answer to the questions raised at the end of
Section~\ref{int}. Still, it is useful to employ again the model used
in these calculations. As we shall see, the model yields a completely
general expression for the conductance and its dependence upon $V_g$
and $\phi$ in the framework of the geometry displayed in
Figure~\ref{fig1}.

Starting point for the study of a case with $R$ leads where $R$ is
integer and $R \geq 2$ is the Landauer--B\"uttiker formula
\begin{equation}
I_r = \sum_{s = 1}^R G_{rs} V_s \ . 
\label{eq0}
\end{equation}
The formula connects the current $I_r$ in lead $r$, $r = 1,\ldots,R$
with the voltages $V_s$ applied to leads $s$. The conductance
coefficients $G_{rs}$ are given by
\begin{equation}
G_{rs} = \frac{e^2}{h} T^{rs} = \frac{e^2}{h} \sum_{ab} \int {\rm
  d}{\cal E} \biggl ( \frac{{\rm d}F({\cal E})} {{\rm d}{\cal E}}
\biggr ) [ |S^{rs}_{ab}({\cal E})|^2 - \delta_{ab} ] \ .
\label{eqg}
\end{equation}
Here $S^{rs}_{ab}({\cal E})$ are the elements of the scattering matrix
$S$ connecting channel $a$ in lead $r$ with channel $b$ in lead $s$ at
an energy ${\cal E}$ of the electron. We have used the terminology of
scattering theory and identified the transverse modes of the electron
in each lead with the channels. The function $F({\cal E})$ is the
Fermi function. We simplify our reasoning by considering very low
temperatures where the integral in Eq.~(\ref{eqg}) can be replaced by
$[ |S^{rs}_{ab}(E_F)|^2 - \delta_{ab} ]$, identifying the Fermi energy
parametrically with the plunger gate voltage $V_g$. We do so because
this brings out the energy dependence of the transmission phase most
clearly. Subsequent averaging over temperature does not change the
essential aspects. We recall that the Landauer--B\"uttiker formula is
restricted to the case of independent electrons. This is the
approximation used throughout the paper.

To proceed, we must introduce a model for the scattering matrix
$S(E_F)$. We consider a geometrical arrangement of the type shown
schematically in Figure~\ref{fig1} without, however, limiting
ourselves to six channels. We emphasize that this geometry differs
from the one considered in Ref.\cite{ent01} where, as mentioned above,
the QD is directly coupled to one of the leads. The electrons move
independently in the two--dimensional area defined by the leads, the
AB interferometer, and the QD. A homogeneous magnetic field is applied
perpendicularly to the plane of Figure~\ref{fig1}.  

The two--dimensional configuration space is divided into disconnected
subspaces defined by the interior of the QD, of the AB interferometer
without the QD, and by each of the leads. The free scattering states
in lead $r$ carry the labels $E$ for energy and $a$ for the channel,
with $c^{r \dagger}_{aE}$ and $c^r_{aE}$ the corresponding creation
and destruction operators. The bound states in the closed AB ring have
energies $\varepsilon_i$, $i = 1,2,\ldots$ and associated operators
$d^{\dagger}_i$ and $d_i$. The QD supports a single bound state with
energy $E_0$ and associated operators $q^{\dagger}$ and $q$. This last
simplification is introduced because we wish to investigate the
behavior of $\delta \phi$ at an {\it isolated} Coulomb--blockade
resonance. At the expense of an increase of the number of indices,
this assumption can easily be removed, see Refs.~\cite{hack96,hack97}.

The single--particle Hamiltonian $H$ is accordingly written as the sum
of two terms,
\begin{equation}
H = H_0 + H_1 \ .
\label{eq1}
\end{equation}
Here, $H_0$ describes free electron motion in each of the disconnected
subspaces,
\begin{equation}
H_0 = \sum_{r,a} \int {\rm d}E \ E c^{r \dagger}_{aE} c^r_{aE} +
\sum_i \varepsilon_i d^{\dagger}_i d_i + E_0 q^{\dagger} q \ . 
\label{eq2}
\end{equation}
The coupling between the various subspaces, and the influence of the
magnetic field are contained in the coupling Hamiltonian
\begin{eqnarray}
H_1 &=& \sum_{r,a,i} \int {\rm d}E \biggl ( W^r_{a i}(E) c^{r
  \dagger}_{a E} d_i + h. c. \biggr ) \nonumber \\
&&\qquad + \sum_{i p} \biggl ( V^p_i q^{\dagger} d_i + h. c. \biggr )
  \ .
\label{eq3}
\end{eqnarray}
The matrix elements $W^r_{a i}(E)$ describe the coupling between
channel $a$ in lead $r$ and the state $i$ in the AB ring. There are no
barriers separating the leads from the AB ring. Therefore, the
coupling to the leads will change the states $i$ into strongly
overlapping resonances. We will accordingly assume later that the
resulting terms depend smoothly on energy $E$. We also assume that on
the scale of the mean level spacing in the AB ring, the energy
dependence of the $W$'s is smooth and, in effect, negligible. The
matrix elements $V^p_i$ describe the coupling between the states $i$
and the state in the QD with energy $E_0$. The upper index $p$ with $p
= 1,2$ distinguishes the two barriers which separate the QD from the
AB ring, see Figure~\ref{fig1}. In our model, the topology of the AB
ring enters via the occurrence of two independent amplitudes for decay
of the state with energy $E_0$ on the QD into each of the states $i$
of the AB ring. Because of gauge invariance, the entire dependence of
$H$ on the applied magnetic field can, without loss of generality, be
put into one of the matrix elements $V$. We accordingly assume that
the amplitudes $W^r_{a i}(E)$ and $V^1$ are real and write $V^2$ in
the form 
\begin{equation}
V^2_i \exp(+i \phi) = (V^{2}_i)^* \exp(-i \phi) = v^{(2)}
\label{eq4}
\end{equation}
where $v^{(2)}$ is real. The phase $\phi$ is given by $2 \pi$ times
the magnetic flux through the AB ring in units of the elementary flux
quantum. Eqs.~(\ref{eq3},\ref{eq4}) imply that the electron picks up
the phase factor $\exp (+i \phi )$ as it leaves the QD through barrier
$2$. Here and likewise in Section~\ref{sca}, we neglect all other
effects that the magnetic field may have on the motion of the
electron, and take account of the Aharanov--Bohm phase only.

The Hamiltonian used in Ref.~\cite{hack96} differs from our $H_0$
in that it also contains the Coulomb interaction between electrons
within the QD in a mean--field approximation. This interaction is
known~\cite{bal99} to be important for the behavior of the phase of
the transmission amplitude through the QD between resonances. It is
not expected, however, to affect this phase in the domain of an
isolated Coulomb--blockade resonance, or the width of such a resonance.

\section{Scattering Matrix}
\label{sca}

One may wonder whether the model defined in Eqs.~(\ref{eq1}) to
(\ref{eq4}) is sufficiently general to give a completely satisfactory
answer to the questions posed at the end of Section~\ref{int}. It is
for this reason that we now derive the form of the scattering matrix
from very general principles. These are unitarity, time--reversal
invariance, the topology of the AB interferometer, gauge invariance,
and the single--level approximation for the passage of electrons
through the QD. As remarked above, the last of these assumptions can
easily be lifted. At the end, it will turn out that the scattering
matrix determined in this way has indeed the same form as the one
calculated from Eqs.~(\ref{eq1}) to (\ref{eq4}).

The total scattering matrix $S$ for the passage of electrons through
the AB interferometer is the sum of two terms. We consider first the
contribution $S^{(0)}$ from that arm of the interferometer which does
not contain the QD. (In the scheme of Figure~\ref{fig1}, the total
scattering matrix $S(E)$ would become equal to $S^{(0)}(E)$ if the
barriers $1$ and $2$ separating the QD from the AB ring were closed).
We neglect the dependence of $S^{(0)}$ on energy over an interval
given by the width of the resonance due to the single level in the QD
in the other arm and, therefore, consider $S^{(0))}$ as independent of
energy. Since $S$ is unitary, and since the contribution from the
other arm vanishes for energies far from the resonance, $S^{(0)}$ must
also be unitary. Moreover, $S^{(0)}$ is not affected by the presence
of the magnetic field. (As in the model of Eqs.~(\ref{eq1}) to
(\ref{eq4}), the entire dependence on the magnetic field will be
contained in the amplitude coupling the QD to the AB ring through
barrier $2$, see Eq.~(\ref{eq13}) below). Hence, $S^{(0)}$ is
time--reversal invariant and, thus, symmetric. As is the case for
every unitary and symmetric matrix, $S^{(0)}$ can be written in the
form~\cite{nis85}
\begin{equation}
S^{(0)} = U U^T \ .
\label{eq11}
\end{equation}
The symbol $T$ denotes the transpose of a matrix, and the matrix $U$
is unitary. It is the product of an orthogonal matrix which
diagonalizes $S^{(0)}$, and of a diagonal matrix. The elements of the
latter have the form $\exp(i \delta)$ where the $\delta$'s are the
eigenphaseshifts of $S^{(0)}$. We now use the more explicit notation
introduced in Eq.~(\ref{eq2}) to write $S^{(0)}$ as
$(S^{(0)})^{rs}_{ab}$, and $U$ as $U^r_{a \alpha}$. With $N_r$ the
number of channels in lead $r$ and $N = \sum_r N_r$ the total number
of channels, the index $\alpha$ runs from $1$ to $N$. The matrix $U$
represents a rotation in the space of channels from the physical
channels $(r,a)$ to the eigenchannels $\alpha$ of $S^{(0)}$.

Using Eq.~(\ref{eq11}), we write the total $S$--matrix $S$ in the form
\begin{equation}
S^{rs}_{ab} = \sum_{\alpha \beta} U^r_{a \alpha} \bigl (
\delta_{\alpha \beta} - i \frac{x_{\alpha \beta}}{E - E_0 + (i/2)
  \Gamma} \bigr ) U^s_{b \beta} \ .
\label{eq12}
\end{equation}
The first term in brackets on the right--hand side of Eq.~(\ref{eq12})
represents $S^{(0)}$ and the second, the contribution of the single
resonance due to the QD. This contribution is written in Breit--Wigner
form. The numerator $x_{\alpha \beta}$ has the form
\begin{eqnarray}
x_{\alpha \beta} &=& z^{(1)}_{\alpha} z^{(1)}_{\beta} +
z^{(2)}_{\alpha} z^{(2)}_{\beta} \nonumber \\
&&\qquad + z^{(1)}_{\alpha} z^{(2)}_{\beta} \exp (i \phi) +
z^{(2)}_{\alpha} z^{(1)}_{\beta} \exp( - i \phi) \ .
\label{eq13}
\end{eqnarray}
The amplitudes $z^p_{\alpha}$ with $p = 1,2$ and $\alpha = 1,\ldots,N$
are the amplitudes for decay of the Breit--Wigner resonance into the
eigenchannels $\alpha$ through the first or the second barrier,
respectively, see Figure~\ref{fig1}. The entire magnetic--field
dependence is contained explicitly in the phase factors. Therefore,
the amplitudes $z^p_{\alpha}$ can be chosen real. The four terms on
the right--hand side of Eq.~(\ref{eq13}) account for the four ways in
which the Breit--Wigner resonance contributes to the scattering
process, see Figure~\ref{fig1}: Formation and decay of the resonance
through barrier one, formation and decay through barrier two,
formation through barrier one and decay through barrier two, and
formation through barrier two and decay through barrier one,
respectively. Thus, the form of Eq.~(\ref{eq13}) accounts for the
topology of the AB ring and for gauge invariance. In writing
Eq.~(\ref{eq13}), we have assumed that passage through the QD is
possible only via intermediate formation of the resonance.

The matrix $S$ in Eq.~(\ref{eq12}) must be unitary. This condition is
met if the total width $\Gamma$ obeys the equation
\begin{eqnarray}
\Gamma &=& \sum_{\alpha} \sum_p (z^{(p)}_{\alpha})^2 + 2 \cos ( \phi )
\sum_{\alpha} z^{(1)}_{\alpha}  z^{(2)}_{\alpha} \nonumber \\
&&\qquad = \sum_{\alpha} | z^{(1)}_{\alpha} + z^{(2)}_{\alpha} \exp( i
\phi ) |^2 \ .
\label{eq14}
\end{eqnarray}
The last form of Eq.~(\ref{eq14}) shows that the amplitude for decay
of the resonance into channel $\alpha$ is the sum of two terms,
$z^{(1)}_{\alpha}$ and $z^{(2)}_{\alpha} \exp ( i \phi )$. Again, this
reflects the topology of the AB interferometer. We note that the
Breit--Wigner term in Eq.~(\ref{eq12}) describes both, single and
multiple passage of the electron through the QD, the latter possibly
in conjunction with multiple turns around the AB ring. This is seen by
expanding the Breit--Wigner denominator in Eq.~(\ref{eq12}) in powers
of $\Gamma$, using Eq.~(\ref{eq14}), and identifying the $n^{\rm th}$
power of pairs of coefficients $z^{(1)}, z^{(2)}$ with the $n$--fold
passage of the electron through the QD. The expansion gives rise,
among many others, for instance to the term
\begin{eqnarray}
&& (\frac{- i}{E - E_0})^n (1/2)^{n-1} \nonumber \\
&& \times (z^{(1)}_{\alpha} [ \sum_{\gamma} z^{(2)}_{\gamma} \exp ( +i
\phi ) z^{(1)}_{\gamma} ]^{n-1} z^{(2)}_{\beta} \exp ( +i \phi ) \ . 
\label{eq30}
\end{eqnarray}
With $(- i)$ the propagator in each of the eigenchannels and $(E -
E_0)^{-1}$ the propagator through the resonance on the QD, this term
describes an electron passing $n$ times through the QD and circling
the AB ring $(n-1)$ times counter--clockwise before leaving the AB
device. (The factor $(1/2)^{(n-1)}$ is a matter of convention).

According to Eq.~(\ref{eq14}), the width $\Gamma$ depends upon the
magnetic flux $\phi$. In the context of the question addressed in the
present paper, this fact is worrysome. Indeed, what is the meaning of
the question ``Does the transmission phase increase by $\pi$ over the
width of the resonance?'' if that width changes with the applied
magnetic field? We now show that the dependence of $\Gamma$ on $\phi$
becomes negligible when the total number $N$ of channels becomes
large, $N \gg 1$. The amplitudes $z^{(p)}_{\alpha}$ are real but may
be positive or negative. For a rough estimate, we assume that the
$z$'s are Gaussian--distributed random variables with zero mean value
and a common variance $z^2$. Then the mean value of $\Gamma$ is easily
seen to be independent of $\phi$ and given by $2 N z^2$. The variance
of $\Gamma$, on the other hand, is given by $[4 N^2 + 4 N (1 + \cos^2
( \phi ))] (z^2)^2$. This establishes our claim: The dependence of
$\Gamma$ on $\cos ( \phi )$ is small of order $1/\sqrt{N}$. To
simplify the discussion, we will assume in the sequel that $\Gamma$ is
independent of $\phi$.

Inspection shows that the matrix $S$ obeys the identity 
\begin{equation}
S ( \phi ) = S^T ( - \phi ) \ .
\label{eq15}
\end{equation}
This equation expresses time--reversal invariance in the presence of a
magnetic field.

For later use, we write, with $f(E)$ real, the Breit--Wigner
denominator in the form
\begin{equation}
\frac{1}{E - E_0 + \frac{i}{2} \Gamma} = f(E) \exp ( i \xi(E) ) \ .
\label{eq16}
\end{equation}
Here, $\xi(E)$ is the resonance phase shift. It increases by $\pi$
over the width $\Gamma$ of the resonance. We recall that we assume
$\Gamma$ to independent of $\phi$. The questions raised at the end of
Section~\ref{int} amount to asking: ``What is the connection between
the transmission phase and the resonance phase shift $\xi(E)$?'' We
will turn to this question presently.

It is useful to introduce the amplitudes
\begin{equation}
\gamma^r_a = \sum_{\alpha} U^r_{a \alpha} z^{(1)}_{\alpha}\ , \
\delta^r_a = \sum_{\alpha} U^r_{a \alpha} z^{(2)}_{\alpha}\ . 
\label{eq17}
\end{equation}
The symbol $\delta^r_a$ should not be confused with the
eigenphaseshift $\delta_{\alpha}$ of $S^{(0)}$. After multiplication
with the matrix $U$, the numerator of the Breit--Wigner term takes the
form
\begin{eqnarray}
\gamma^r_a \gamma^s_b &+& \delta^r_a \delta^s_b + [\gamma^r_a \delta^s_b
+ \delta^r_a \gamma^s_b] \cos ( \phi ) \nonumber \\
&&\qquad + i [\gamma^r_a \delta^s_b - \delta^r_a \gamma^s_b] \sin (
\phi ) \ .
\label{eq18}
\end{eqnarray}
The total scattering matrix is given by
\begin{eqnarray}
S^{rs}_{ab}(E) &=& (S^{(0)})^{rs}_{ab} - i f(E) \exp ( i \xi(E) ) [
\gamma^r_a \gamma^s_b + \delta^r_a \delta^s_b \nonumber \\
&&+ [\gamma^r_a \delta^s_b + \delta^r_a \gamma^s_b] \cos (
\phi ) \nonumber \\
&&\qquad + i [\gamma^r_a \delta^s_b - \delta^r_a \gamma^s_b] \sin (
\phi ) ] \ . 
\label{eq18a}
\end{eqnarray}
The width $\Gamma$ can also be expressed in terms of the amplitudes
$\gamma^r_a$ and $\delta^r_a$,
\begin{equation}
\Gamma = \sum_{ra} |\gamma^r_a + \delta^r_a \exp ( i \phi ) |^2 \ .
\label{eq18b}
\end{equation}

As announced above, we have shown that the scattering matrix $S$ can
indeed be constructed from the requirements of unitarity,
time--reversal invariance, the topology of the AB interferometer,
gauge invariance, and the single--level approximation. Explicit
construction of the scattering matrix from the Hamiltonian formulated
in Eqs.~(\ref{eq1}) through (\ref{eq4}) as done in Ref.~\cite{hack96}
yields an expression which is identical in form to our $S$ in
Eq.~(\ref{eq18a}). This shows that our formal construction possesses a
dynamical content. Conversely, this result shows that our model
Hamiltonian in Eqs.~(\ref{eq1}) to (\ref{eq4}) leads to the most
general form of the scattering matrix which is consistent with the
requirements just mentioned. We recall that our construction is
strictly based upon a single--particle picture and does not account
for interactions between electrons beyond the mean--field
approximation.

\section{The Transmission Phase}
\label{tra}

Equipped with an explicit expression for $S$, we return to the
transmission phase. It should first be noted that different
experiments may determine different combinations of the conductance
coefficients $G_{rs}$ introduced in Eq.~(\ref{eq0}). As shown in
Ref.~\cite{hack97}, the relevant quantity for the experiment of
Schuster {\it et al.}~\cite{sch97} is $T^{41} /(T^{44} - N_4)$. Here,
the indices $1$ and $4$ label the source and the collector,
respectively, for the electrons in the six--lead experiment. We will
show presently that $T^{rr}$ with $r = 1,\ldots,R$ is an even function
of the phase $\phi$ and, therefore, depends only upon $\cos ( \phi)$.
A non--trivial dependence on $\phi$ involving both $\cos (\phi)$ and
$\sin (\phi)$ and, thus, a trigonometric dependence on $(\phi \pm
\xi(E_F))$, arises only from the terms $T^{rs}$ with $r \neq s$. 
Without loss of generality we, therefore, focus attention on $T^{12}$
and, thus, on $\sum_{ab} | S^{12}_{ab}(E_F) |^2$, see Eq.~(\ref{eq0}).
This quantity is expected to display a non--trivial dependence on
$\phi$. We expect that the transmission phase $\delta \phi$ increases
by $\pi$ as the Fermi energy sweeps the Coulomb--blockade resonance. We
ask how this increase depends on the width $\Gamma$ of the resonance
and on the number $R$ of leads.

It is useful to address these questions by using the unitarity
relation. We write
\begin{equation}
\sum_{ab} | S^{12}_{ab} |^2 = N_1 - \sum_{ab} | S^{11}_{ab} |^2 -
\sum_{s \geq 3} \sum_{ab} | S^{1s}_{ab} |^2 \ .
\label{eq19}
\end{equation}
The advantage of Eq.~(\ref{eq19}) is that the sum over $s$ vanishes
when there are only two leads. Thus, the influence of the number of
leads is made explicit. We now discuss the dependence of the terms on
the right--hand side of Eq.~(\ref{eq19}) on the resonance phase shift
$\xi(E)$.

Each of the terms $\sum_{ab} | S^{1s}_{ab} |^2$ with $s =
1,3,4,\ldots$ in Eq.~(\ref{eq19}) is the sum of three contributions,
involving $|S^{(0)}|^2$, the square of the Breit--Wigner contribution,
and the interference term between $S^{(0)}$ and the Breit--Wigner
term. The contributions from $\sum_{ab} | (S^{(0)})^{1s}_{ab} |^2$ are
independent of both, energy and magnetic field and, therefore, without
interest. These terms only provide a smooth background. The squares of
the Breit--Wigner terms are each proportional to $f^2(E_F)$ and are
independent of the resonance phase shift $\xi(E_F)$. This is expected.
Each such term depends on the magnetic flux $\phi$ in two distinct
ways. The squares of the terms in Eq.~(\ref{eq18a}) involving either
$\cos ( \phi )$ or $\sin ( \phi )$, and the product of these two terms
yield a dependence on $\phi$ which is periodic in $\phi$ with period
$\pi$. Such terms can easily be distinguished experimentally from
terms which are periodic in $\phi$ with period $2 \pi$. As we shall
see, it is some of these latter terms which carry the resonance phase
shift $\xi(E_F)$. Therefore, we confine attention to terms of this
latter type. A contribution of this type arises from the square of the
Breit--Wigner term via the interference of that part of the resonance
amplitude which is independent of $\phi$ with the terms proportional
to either $\cos ( \phi )$ or $\sin ( \phi )$. The sum of all such
contributions (from values of $s = 1$ and $s = 3,4,\ldots,R$) has the
form
\begin{equation}
f^2(E_F) A \cos ( \phi + \alpha_0 )
\label{eq19a}
\end{equation}
where $A$ and $\alpha_0$ are constants wich depend on $R$ but not on
$E_F$ or $\phi$. We note that the constant $A$ is of fourth order in
the decay amplitudes $\gamma^r_a$ and $\delta^r_a$. Isolated
resonances in quantum dots require high barriers, i.e., small values
of these decay amplitudes. Therefore, the term~(\ref{eq19a}) may be
negligible. We continue the discussion under this assumption but note
that there is no problem in taking account of this term if the need
arises.

We turn to the interference terms. We first address the case $s = 1$
which differs from the cases with $s \geq 3$. We observe that
$(S^{(0)})^{11}_{ab}$ is even in $a,b$. Therefore, multiplication of
$S^{(0)}$ with the Breit--Wigner term and summation over $a$ and $b$
will cancel those parts of the Breit--Wigner term which are odd under
exchange of $a$ and $b$. Inspection of Eq.~(\ref{eq18}) shows that
these are the terms proportional to $\sin ( \phi )$. As a consequence,
the interference term for $s = 1$ is even in $\phi$ and a function of
$\cos ( \phi )$ only. Explicit calculation shows that the term
proportional to $\cos ( \phi )$ can be written in the form
\begin{equation}
4 f(E_F) x^{(1)} \cos ( \phi ) \sin ( \xi(E_F) + \zeta^{(1)} ) \ .
\label{eq20}
\end{equation}
Here, $x^{(1)}$ and $\zeta^{(1)}$ are independent of energy and
explicitly given by $x^{(1)} \exp ( i \zeta^{(1)} ) = \sum_{ab}
\gamma^{(1)}_a \delta^{(1)}_b ((S^{(0)})^{11}_{ab})^*$. We observe
that as a function of energy, $\sin ( \xi(E_F) + \zeta )$ always has a
zero close to the resonance energy $E_0$. When only two channels are
open, the entire dependence on $\phi$ which has period $2 \pi$ resides
in this term. The term does not display the resonance phase shift
except through the zero near $E_0$. It is symmetric in $\phi$ about
$\phi = 0$. These facts are well known~\cite{bue86}, of course, and
are reproduced here for completeness only. The form~(\ref{eq20}) of
the interference term was responsible for the failure of the
experiment in Ref.~\cite{yac95} to measure the transmission phase.

For the interference terms with $s \geq 3$, the matrix
$(S^{(0)})^{1s}_{ab}$ is not symmetric in $a,b$. (It is symmetric only
with respect to the simultaneous interchange of $1,s$ {\it and}
$a,b$). Therefore, the terms proportional to $\sin ( \phi )$ in
Eq.~(\ref{eq18a}) do not cancel, and the interference terms acquire a
genuine joint dependence on both, the resonance phase shift $\xi(E_F)$
and the phase $\phi$ of the magnetic flux. Proceeding as in the
previous paragraph, we introduce the constants $x \exp ( i \zeta ) =
\sum_{s \geq 3} \sum_{ab} \gamma^{(1)}_a \delta^{(s)}_b
((S^{(0)})^{1s}_{ab})^*$ and $y \exp ( i \theta ) = \sum_{s \geq 3}
\sum_{ab} \gamma^{(s)}_a \delta^{(1)}_b ((S^{(0)})^{1s}_{ab})^*$. The
$\phi$--dependent part of the sum of the interference terms with $s
\geq 3$ takes the form
\begin{equation}
2 f(E_F) [ x \sin (\phi + \xi(E_F) + \zeta) + y \sin(-\phi + \xi(E_F)
+ \theta) ] \ .
\label{eq21}
\end{equation}
This expression depends on $\xi(E_F)$ in the expected way.

We are now in a position to answer the questions raised at the end of
Section~\ref{int}. Whenever the total number $N$ of channels coupled
to the AB device is sufficiently large, the resonance width $\Gamma$
becomes independent of magnetic flux. This property can be checked
experimentally. It is only in this limit that the statement ``The
resonance phase shift increases by $\pi$ over the width of the
resonance'' acquires its full meaning. The limit $N \gg 1$ may, of
course, be realised even when the number $R$ of leads is small. We
turn to the behavior of the transmission phase. We have shown that
there are terms proportional to $f^2(E_f)$ which depend upon $\cos (
2 \phi )$ but not on $\xi(E_F)$. The form of these terms was discussed
above. For a quantum dot with high barriers, it is expected that these
terms are small. The terms periodic in $\phi$ with period $2 \pi$ are
listed in Eqs.~(\ref{eq19a}) to (\ref{eq21}). For a quantum dot with
high barriers, we expect that the contribution~(\ref{eq19a}) is small.
We focus attention on the remaining terms. These depend on the value
of $R$. For $R = 2$, the phase dependence is given by the
term~(\ref{eq20}). This term is even in the magnetic flux $\phi$ and
has a zero close to the resonance energy $E_0$. It does not, however,
display the smooth increase of the resonance phase phase shift
$\xi(E_F)$ over the width $\Gamma$ of the resonance. If, on the other
hand, the number of leads $R$ is large compared to unity, then it is
reasonable to expect that the terms in Eq.~(\ref{eq21}) are large
compared to the term in Eq.~(\ref{eq20}). This is because the number
of contributions to the terms in Eq.~(\ref{eq21}) is proportional to
$R - 2$. In this case, the transmission phase faithfully reflects the
energy dependence of the resonance phase shift $\xi(E_F)$. Deviations
from this limit are of order $1/(R - 2)$. As we gradually turn off the
coupling to the leads $s$ with $s \geq 3$, the terms~(\ref{eq21})
gradually vanish. Nevertheless, it is possible --- within experimental
uncertainties that become ever more significant as the
terms~(\ref{eq21}) become smaller --- even in this case to determine
the resonance phase shift $\xi(E_F)$ from the data on the transmission
phase. We propose the following procedure. Add formulas~(\ref{eq20})
and (\ref{eq21}) and fit the resulting expression to the data. This
should allow a precise determination of $\xi(E_F)$ and of $\Gamma$
also in cases where the coupling to the leads $s$ with $s \geq 3$ is
small. This statement holds with the proviso that the number of
channels must be large enough to allow us to consider the total width
$\Gamma$ as independent of $\phi$. Whenever the coupling to the leads
$s$ with $s \geq 3 $ is small, the energy dependence of the
transmission phase is quite different from that of the resonance phase
shift. Nevertheless, the transmission phase $\delta \phi$ does reflect
the energy dependence of the resonance phase shift $\xi$ whenever the
number of leads is larger than two. In particular, this energy
dependence is governed by the total width $\Gamma$.

In conclusion, we have seen that in a theoretical model which is more
closely patterned after the geometry of Figure~\ref{fig1} than is the
model of Ref.\cite{ent01}, the transmission phase does reflect the
value of the total width. This statement applies whenever the number
of leads exceeds the value two, and whenever the total number of
channels is large compared to one. Both conditions are expected to be
met in the experiment of Schuster {\it et al.}~\cite{sch97}.

{\it Acknowledgment.} I am grateful to O. Entin--Wohlman, A. Aharony,
and Y. Imry for informative discussions which stimulated the present
investigation. I thank O. Entin--Wohlman for a reading of the paper,
and for useful comments. This work was started when I was visiting the
ITP at UCSB. The visit was supported by the NSF under contract number
PHY99-07949.

\end{multicols}

\end{document}